# VERIFICATION OF BETHE-BLOCH FORMULA USING GEANT4 TOOLKIT

# C.1 Bremsstrahlung Radiation loss

# C2. Cherenkov Radiation

Vanshaj Kerni

Undergraduate, Departement of Physics, Indian Institute of Technology Roorkee, Roorkee, 247667, Uttrakhand

Dr. Jyothsna Rani Komaragiri

Center for High Energy Physics, Indian Institute of Science, Bangalore, 560012, Karnataka

# Table of Contents

A : Density and Shell Corrections

B: Adiabiatic Invariance

C: High energy losses

Source

# VERIFICATION OF BETHE-BLOCH FORMULA USING GEANT4 TOOLKIT

Vanshaj Kerni

Undergraduate, Departement of Physics, Indian Institute of Technology Roorkee, Roorkee, 247667, Uttrakhand

Dr. Jyothsna Rani Komaragiri

Center for High Energy Physics, Indian Institute of Science, Bangalore, 560012, Karnataka

## Abstract

Understanding of the physcial world is a continuous process of trial and error. It involves the development of the theoretical background for the process we wish to understand and then use the necessary tools to try and find out the accuracy of the theory. One such process is the interaction of particles passing through matter and their energy loss. Testing the energy loss experimentally required the development of a tool that could be used to simulate and test the theory with actual data. GEANT4 is the tool that was developed. A multi-purpose toolkit used by researchers, scientists to explore the nature of collisions and energy loss of particles. A Monte-Carlo simulation toolkit, used for simulating the passage of particles through various materials. Geant4 is widely used in fields where particle interaction is required ranging from high energy physics, to medical physics, space physics research. The physics component offers a wide range of simulations and tools to study the passage and processes involved. Electromagnetic component is one of the components of Geant4 which is used to verify the Bethe-Bloch energy loss formula for charged particles developed mostly by Hans Bethe. This study includes the work done to verify Bethe-Bloch using GEANT4 and study the extent of it validity.

**Keywords or phrases**: *GEANT4, interaction, energy loss, scattering, collision, charged, particle*

## Abbreviations

Abbreviations

| | |
|---|---|
| A | Mass number |
| Ze | Charge of material (target) |
| ze | Charge of incident particle |
| E | Energy of the incident particle |
| QM | Quantum Mechanics |

# 1 INTRODUCTION

## 1.1 Background

In physical science, a particle is described as a **localized object or entity** which can be described by physical parameters like *volume, density, mass.* These particles can vary from a small pebble to a subatomic particle that makes the atoms to the fundamental particles that are mediator of fundamental forces of nature. These particles are the fundamental reasons of our existence and tempting to gather information regarding them is very intutive human nature.

The early development of the modern physics has been marked with the discovery of radiations, namely α and β radiations. In the early stages physicist, *Ernest Rutherford,* were majorly involved in understanding the nature of these radiations because of different absorptions in matter. He started working on the theoretical ideas behind the interaction of these radiation with matter. Physicist, J.J Thomson, played a crucial role in defining the link of these radiation with the particles that we know today, mainly, *protons* and *electrons.* Thomson, showed that these radiations were indeed particles. Rutherford, through his famous α-scattering experiment which produced scintillations, invoked the problems like *collisions, scattering* of the particles that made the radiations, to explain the scintillation phenomenon[1]. Bohr took the ideas further, developing a theory that could explain the problems put forward by Rutherford. He developed a classical relation for the energy loss of the particles (α & β) in matter. With the improvement in the experimental precision, the classical theory went under various modification, major being by *Hans-Bethe,* who calculated the energy loss using the Quantum Mechanics. Modern energy loss theory has been tested and improved to counter various difficulties in different energy ranges.

### 1.1.1 Why it is important to understand these interactions

Radiations, in the modern science has been a part of many fields, including *High Energy physics, Medical science, Radiation treatments* etc. Understanding, how particles that makes these radiations, such as in *proton thearpy*, *Fast neutron therapy,* interact with matter is of crucial importance both for medical field and for fundamental sciences. Also, for new age students, persuing in these areas, understanding the theory that explains these interactions (in terms of energy loss) and the verification techniques are also of same importance.

## 1.2 Statement of the Problem

To understand the energy loss theory for charged particles moving through matter, mainly the Bethe-Bloch relation and use the GEANT4 toolkit to simulate the passage of charged particles and verify the theory.

## 1.3 Objectives of the Study

- To learn about the theory of energy loss of particles in matter both the classical and QM.
- To understand and work on GEANT4 software to simulate the passage of particles.
- To make plots of the simulated data of GEANT4.
- To verify it with the data using *NIST (National Institute of Standards and Technology) Standard Reference Database 124.*

## 1.4 Scope

The detailed study of the energy loss of charged particles moving through matter has been done in energy range of MeV's to TeV's. The lower energy contribution to the overall energy loss has not been studied under this study. However the verifivation will be done in the energy range of keV's to MeV's.

# 2 BACKGROUND REVIEW

## 2.1 Early concepts and definations

### 2.1.1 Relativistic Force

In classical mechanics, the concept of force has been defined by Newton's second law of motion, which states "the rate change of momentum is force". Mathematically it can be written as

$$\mathbf{F} = \frac{\partial \mathbf{p}}{\partial t} \ldots p = momentum\ of\ particle \tag{i}$$

Using relativistic momentum for subatomic particles, we can calculate the force acting on a particle moving at relativistic speeds. This is important as subatomic particles, having low mass have high momenta

$$\mathbf{p} = m\gamma\mathbf{v} \tag{ii}$$

$$\mathbf{F}_{rel} = m_0\gamma\mathbf{a_t} + m_0\gamma^3(\mathbf{a_t}\cdot\beta)\beta \ldots \mathbf{a_t} = \frac{\partial \mathbf{v}}{\partial t} \tag{iii}$$

$$\mathbf{F_{rel}} = m_0\gamma\mathbf{a_t} + m_0\gamma^3(\mathbf{a_t}\cdot\beta)\beta \ldots \mathbf{a_t} = \frac{\partial \mathbf{v}}{\partial t} \tag{iv}$$

Thus, the relativistic force contains two contributions, one in direction of $\mathbf{a_t}$ and other along $\beta$.

### 2.1.2 Energy-Momentum Relation

Since particles such as *protons, electrons, muons* etc have low masses compared to atoms as whole, their momentum is usually high with velocity in the relativistic regions. Therefore we are required to use the relativistic energy relations. For a particle of mass 'm' and velocity 'v' and momentum 'p', moving in a free space, the total energy is given by

$$E = \sqrt{p^2c^2 + m^2c^4} \tag{v}$$

Using momentum relation $|\mathbf{p}| = m\gamma|\mathbf{v}|$, the kinetic energy of the particle can be written as

$$E_{kin} = mc^2(\gamma - 1) \tag{vi}$$

Therefore if we know the kinetic energy and momentum loss we can find out the mass of the particle.

## 2.1.3 Charged particle and Magnetic Field

Magnetic field becomes a great tool when we deal with charged particles. This is beacuse magnetic field has the property to change the direction of the charged particles moving through the space. The force that acts on the charged particle moving with velocity $\mathbf{v}$ through the space of magnetic field $\mathbf{B}$ is given by *Lorentz Force* relation as

$$\mathbf{F_m} = q\left(\mathbf{v} \times \mathbf{B}\right) \tag{vii}$$

Magnetic field is to find about the momentum of the charged particle moving in magnetic field. Let a charged particle of mass 'm', momentum 'p' moves through a unifrom magnetic field 'B', by measuring the radius 'r' of the circle in which it rotates, momentum is given by

$$\frac{\mathbf{p}\cdot\mathbf{p}}{r^2\gamma^2 Ze}(-\mathbf{r}) = \mathbf{p} \times \mathbf{B} \tag{viii}$$

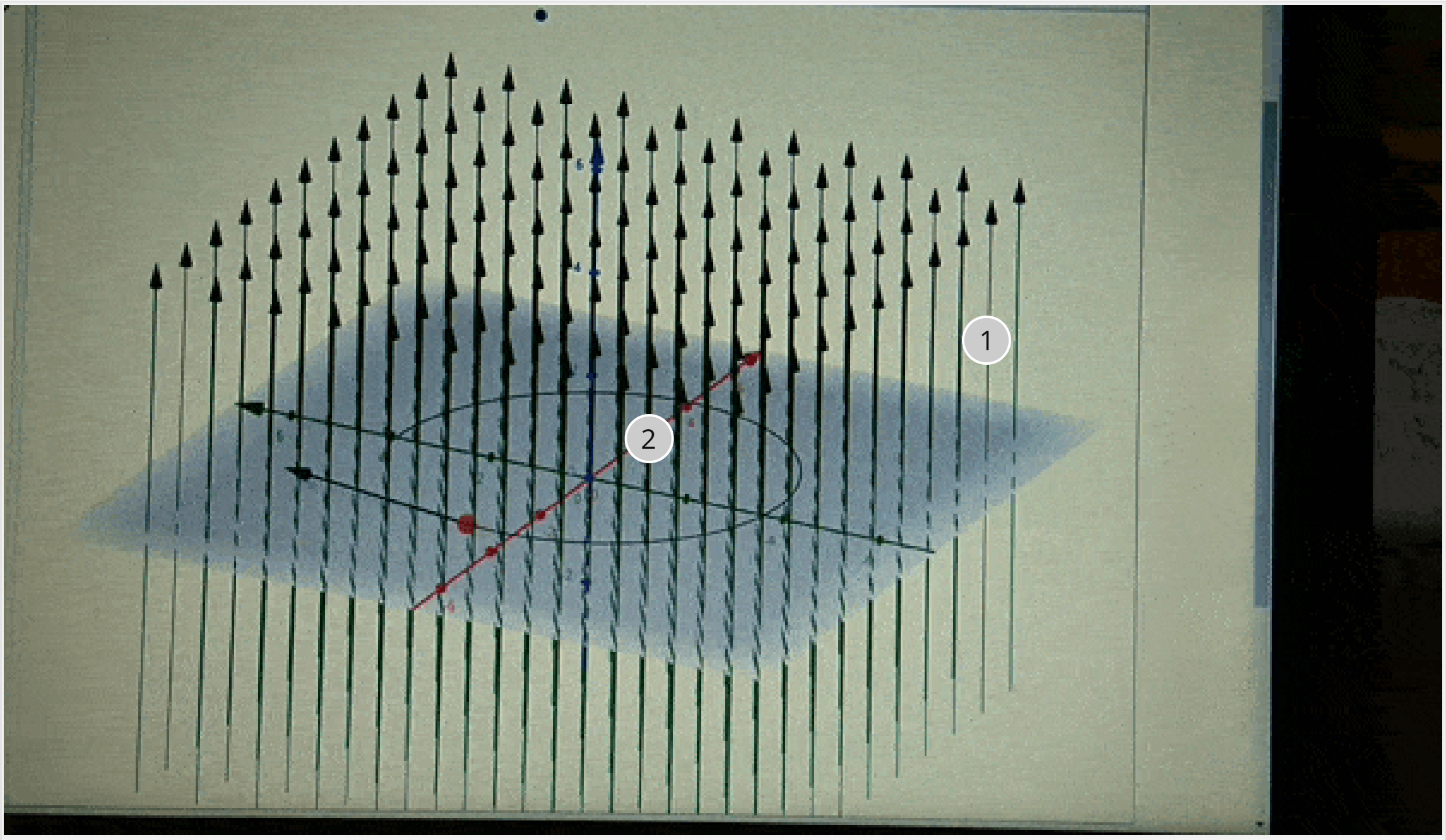

1 magnetic field lines 2 raidus 'r'

Fig 1 charged particle with velocity v moving in magnetic field of strength B pointing in the positive z direction

## 2.1.4 Cross section[2]

The interactions or collision of particles with other particles is defined in terms of *cross section* parameter. It defines the interaction probability, in other terms it defines the extent to which collision of particles takes place. It may be calculated if the collision parameters are known. To give a formal mathematical description, let us assume that a uniform beam of particles with intensity 'I' is incident to a small area such that the diameter of the beam is larger than the area of incidence, then we can define the flux **F** of the incident beam. Also, if we assume that the number of particles scattered into a small solid angle **dΩ** is **dN**, although the number is dependent upon the impact parameters, but over large time it tends to a fixed quantity per unit solid angle, thus the differential cross section **dσ** is defined as

$$\frac{d\sigma}{d\Omega}(E,\Omega) = \frac{1}{F}\frac{dN_s}{d\Omega} \ldots N_s{=}the\ number\ of\ scattered\ particles \tag{ix}$$

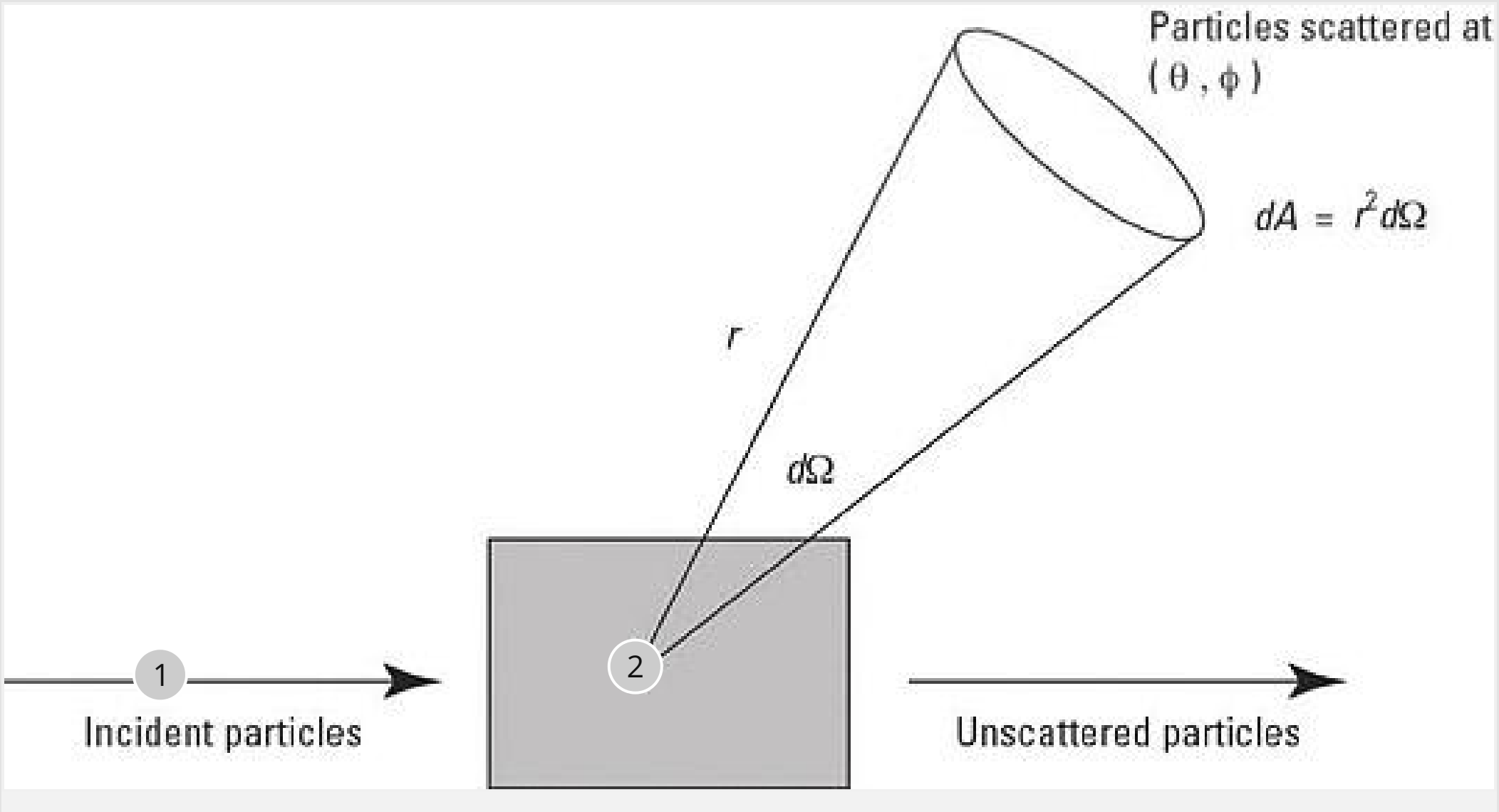

1 Flux F 2 small area

Fig 2 figure showing incident beam of particles that is scattered into a small solid angle

The total cross section can therefore be calculated as

$$\sigma_s(E) = \int \frac{d\sigma(\Omega)}{d\Omega} d\Omega \tag{x}$$

### 2.1.5 Range of particles

As the energy loss by charged particles moving through matter is not continuous, thus the range is accurately defined in a statistical way. However, if we assume that the energy loss is indeed continuous, then range can be defined as the maximum distance travelled (in approximately straight line) by the charged particles in the material. It can be written as

$$\mathcal{R} = \int_0^x dx \ldots x = maximum\ distance\ travelled \quad \text{( xi )}$$

### 2.1.6 Energy loss of charged particles

Charged particles moving through matter interacts with the constituents of matter in a number of ways. In general, most of these interactions can be reduced down to two events:

- Energy loss by the charged particles on being scattered by the particles of matter.
- Change in the direction of the particles with respect to the incident direction.

Both of these events are the result of scattering of the incident particles. This scattering of the incident particle, results in energy loss by the particles as they travel through matter. The energy loss depends upon the charge particle and the interaction mechanism. The most common energy loss is the *energy loss via collision* with atoms which is termed as *ionization energy loss.* Other form of energy loss includes the radiative energy loss, which include *Cherenkov radiation loss, Bremsstrahlung radiation loss, Transition Radiation loss* and low energy losses.

### 2.1.7 Scattering: Elastic-Inelastic, Multiple Scattering and Backscattering

"A general physical process where quanta of some form, such as light, sound, or moving particles, are forced to deviate from a straight trajectory by localized non-uniformities in the medium through which they pass."

In simplar terms, scattering, in particle physics, can be defined as the change in the direction of motion of the incident particle with respect to the incident direction. The reason for scattering of particles in matter are the atoms present at certain locations and the potential developed by those atoms that tends to deviate the charge particle from its original direction.

*Elastic scattering*, may be referred to as elastic collision in classical sence mainly occurs with nucleus of the atoms of material. Elastic collision conservs the kinetic energy and thus not that responsible for the energy loss of the particles. The scattering cross section for non-relativistic particles is given by *Rutherford* formula

$$\frac{d\sigma_s}{d\Omega} = \frac{1}{4} Z_1^2 Z_2^2 r_e^2 \left(\frac{mc}{\beta p}\right)^2 \frac{1}{(sin(\theta/2))^4} \tag{xii}$$

The elastic cross-section for relativistic scattering of two particles(e-'s) is given by Moller in center of mass frame by

$$\frac{d\sigma}{d\Omega}_{moller} = \frac{\alpha^2}{4s}\left[\frac{10+4x+2x^2}{(1-x)^2} + \frac{10-4x+2x^2}{(1+x)^2} + \frac{16}{(1-x)(1+x)}\right] \ldots s{=}total\ c.o.m\ energy\ \ ;\ x{=}cos(\theta) \tag{a}$$

*In-elastic scattering* on the other hand can be regarded as in-elastic collisions in classical sense. In in-elastic collisions, energy is transferred to the atoms of the material and therefore it is the major contributor to the energy loss of the incident particle. In-elastic collisions takes place with the electrons of the atoms and thus, although the energy transferred during a single collision doesnot alter the incident energy, but the number of electrons in the material is very large, thus contributing to a sufficient energy loss. This energy loss is *statistical* in nature, occring with some probabilities. However, this can be averaged over length; considering that the length under consideration should be more than the seperation between atoms providing more collision per length that can be averaged out. Thus an average energy loss can be calculated.[2]

In reality, incident particle interacts with many number of atoms and electrons in the material and is scattered number of times. In literature, if the total number of individual in-elastic scattering 'N' is more than 20, then this comes under *Multiple Scattering*. A particle passing through a material may undergo single of multiple scattering. In general, it is the combination of both that gives the total energy loss of particles. Multipe scattering is the reason for various effects such as *Backscattering, Channeling.*

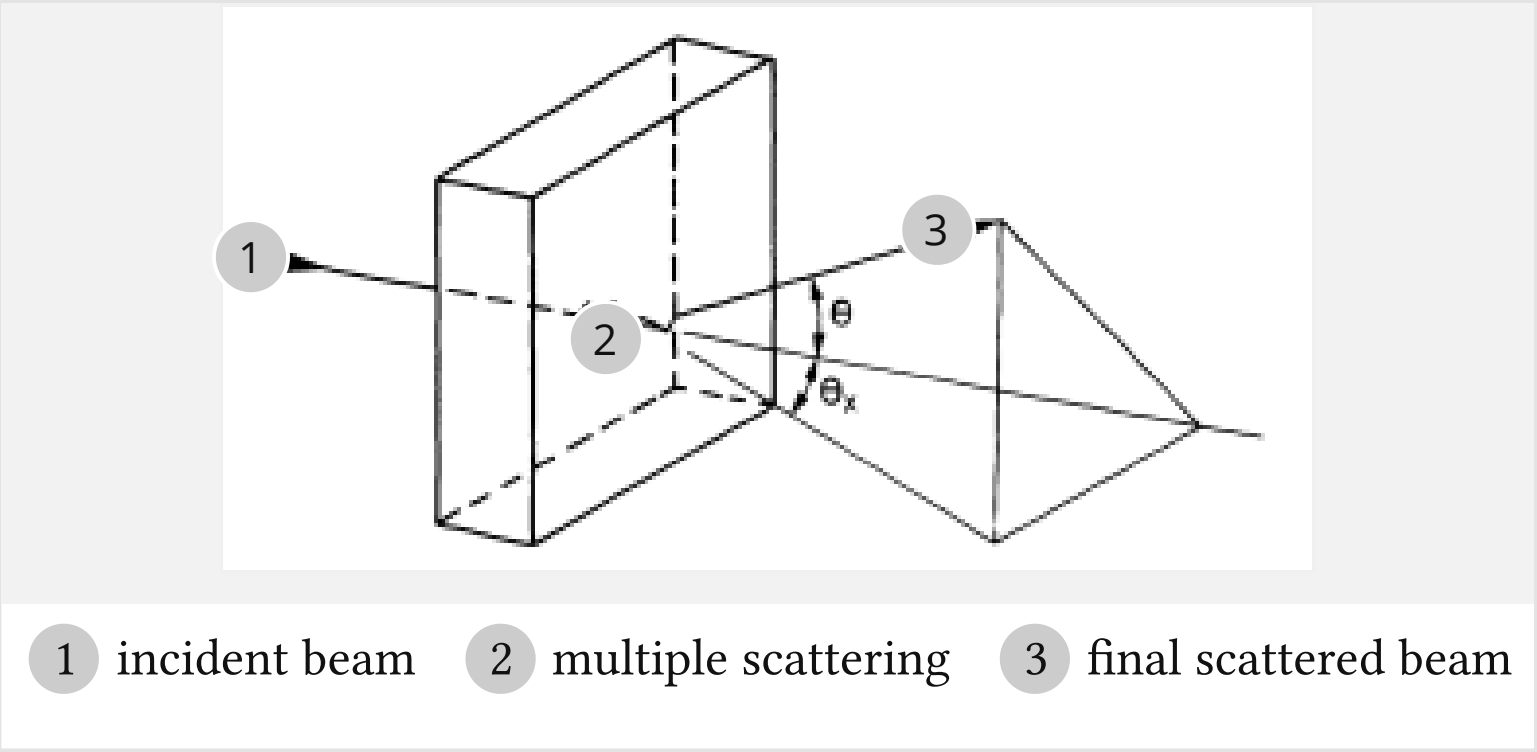

Fig 3 multiple scattering of charged particle moving through matter, deflected from incident direction

Backscattering is the result of multiple scattering of low energy particles. This mainly occurs for slow electrons, $\beta \leq 0.05$, due to their low mass. This may result into the ejection of electron from the same face where it was incident.

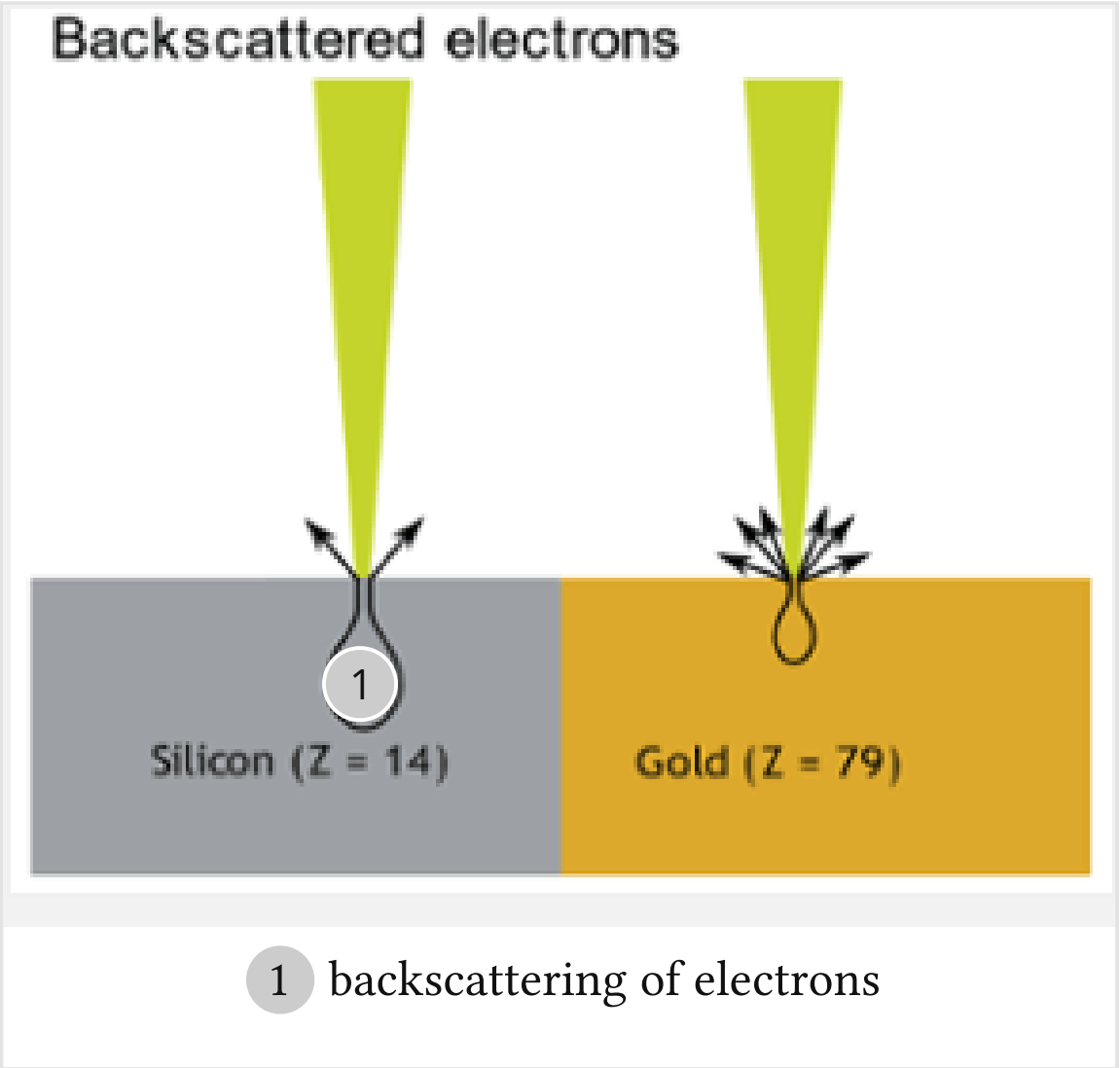

Fig 4 Backscattering of electrons, the extent depends upon the interaction with atoms and electrons, therefore the fraction is more for Gold than Silicon.

## 2.2 Energy loss relations

As mentioned above the energy loss can be estimated as an average energy loss of particles. This was first calculated by *Bohr* using classical methods and is termed as *Bohr classical formula* and then using quantum mechanics by *Bethe.*

### 2.2.1 Bohr classical energy loss formula

Assuming a charged particle (**ze**) moving along x-direction through a medium with electron density '**η**'. Let us assume that it intracts with a single electron (at rest) of the material at a seperaction of 'b' along perpendicular direction as shown

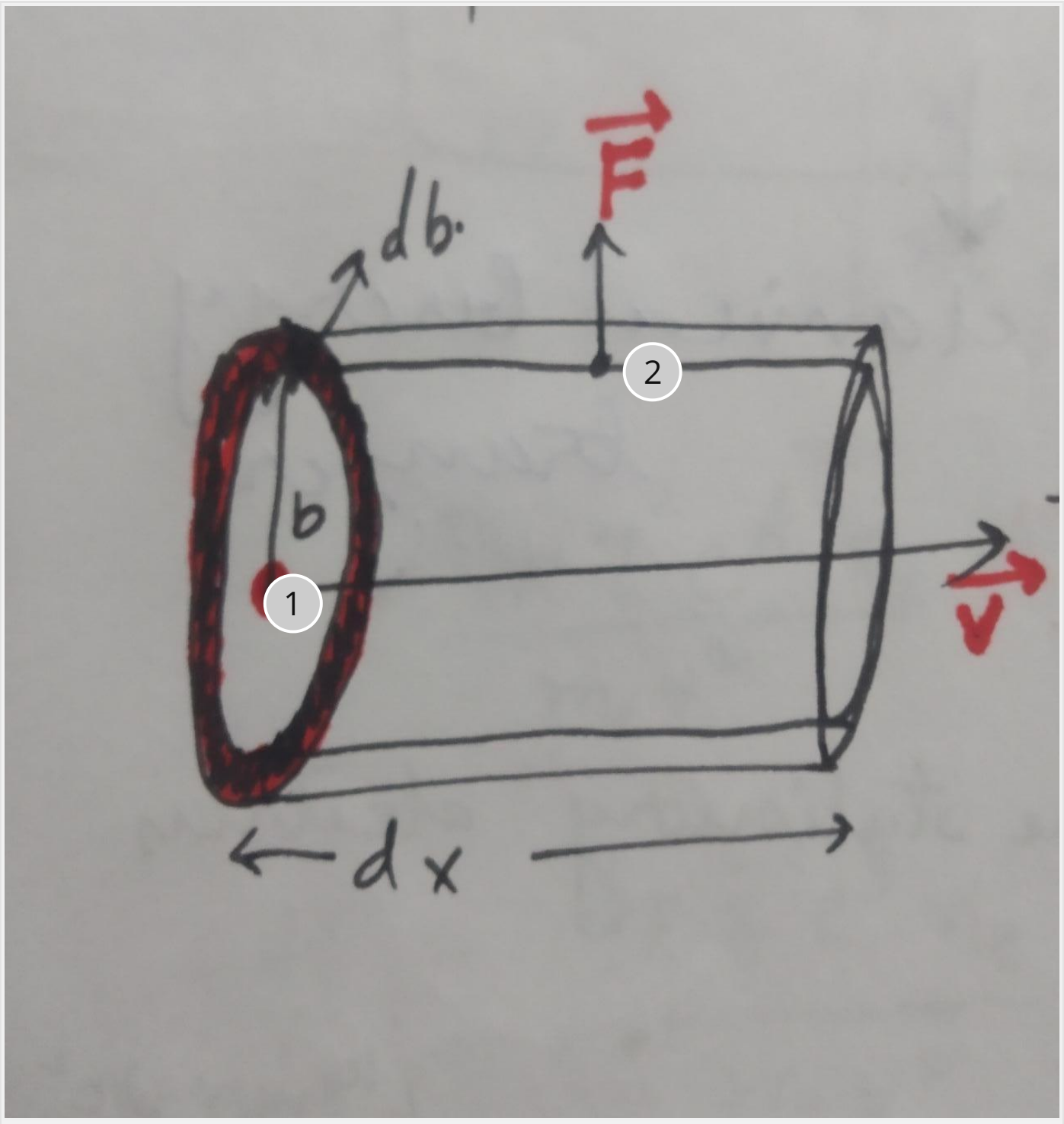

1 charged particle 2 single electron

Fig 5 charged particle with velocity v moving along x direction intracts with a single electron at distance of b along the vertical direction, initially at rest.

Under these assumptions, the momentum transferred to the electron can be calculated by

$$\Delta p = \int (F_{per} + F_{||})dt \implies \int \frac{(F_{per}+F_{||})}{\beta c} dx \ldots_{v=\beta c} \quad \text{( xiii )}$$

It can be easily seen that the effective transfer takes place along the vertical direction therefore;

$$\Delta p_{||} = \int \frac{F_{||}}{\beta c} dx = 0$$

( xiv )

For a differential distance 'dx' travelled, and assuming that the cylinder is infintely long, we can use *Gauss Law* to calculate the perpendicular electric field $E_{per}$ of the charged particle.

$$\int E_{per}\ 2\pi b\ dx = 4\pi z e$$

( xv )

Substituting Equation xiv and Equation xv in Equation xiii, we get

$$\Delta p_{per} = \frac{2ze^2}{\beta c b}$$

( xvi )

Using the kinetic energy - momentum relation of $p^2/2m$, the energy $E_{kin}(b)$transferred to one electron is given by

$$E_{kin}(b) = \frac{\Delta p^2}{2m} \implies \frac{2z^2 e^4}{\beta^2 b^2 c^2 m_e}$$

( xvii )

Thus the total energy transferred to electrons in the width db is given by

$$dE = E_{kin}(b)\ dN \implies E_{kin}(b)\ 2\pi b\eta\ db\ dx$$

( xviii )

Thus the total energy loss per unit length is given by

$$-\left\langle \frac{dE}{dx} \right\rangle = \frac{4\pi z^2 e^2 \eta}{m_e v^2} \int_{bmin}^{bmax} \frac{db}{b}$$

( xix )

Using Equation xvii and momentum conservation with the assumption that $m_{incident\ particle} >> m_e$, bmin can be calculated as

$$b_{min} = \frac{ze^2}{m_e \gamma v^2}$$

( xx )

and using *adiabiatic invariance* bmax can be estimated to $b_{max} = \frac{\gamma v}{\nu'} \ldots \nu' = average\ frequency$. Substituting values of bmax and bmin to Equation xviii, the classical energy loss relation can be found out to be

$$-\left\langle \frac{dE}{dx} \right\rangle = \frac{4\pi z^2 e^2 \eta}{m_e v^2}\ ln\left(\frac{\gamma^2 v^3 m_e}{z e^2 \nu'}\right) \ldots Bohr\ classical\ energy\ loss\ relation \quad \text{( xxi )}$$

## Problems with classical relation

With the development of QM, classical relation showed various problems.

- The relation was not able to predict the energy loss for low mass particles having mass less than the alpha particle.
- It deviates from the energy losses at higher energies.
- It was not able to predict the energy loss of electrons and positrons.

## 2.2.2 Bethe-Bloch energy loss relation

The QM derivation of the energy loss of particles in matter was done by Bethe and contributions from others.[2] Bethe calculated the energy loss formula, taking the relativistic effects into consideration and used the momentum instead of impact parameter 'b' to avoid the problem of figuring the values of b, that Bohr did. Using these, the formula for the energy loss that is used for studies is given by

$$-\left\langle \frac{dE}{dx} \right\rangle = 2\pi N_a r_e^2 m_e c^2 \rho \frac{Z z^2}{A\beta^2}\ \left[ln\left(\frac{2 m_e c^2 \gamma^2 \beta^2 E_{max}}{I^2}\right) - 2\beta^2 - \delta - \frac{C}{Z}\right] \quad \text{( xxii )}$$

where, $E_{max} = \frac{2 m_e c^2 \kappa^2}{1+2s\sqrt{1+\kappa^2+s^2}} \ldots \kappa = \beta\gamma; s = \frac{m_e}{M}$. $E_{max}$ is the maximum energy transfer to an electron in a single collision and I is the *mean ionization potential* of the atom of the elecctron.

This is the relation that is being used for calcultaing the energy loss of charged particles passing through the material. The formula Bethe derived didnot contain the last two correction terms namely, *Density correction (δ)* and *Shell correction (C)*. Density correction is dependent on the incident energy and is given by $\delta = ln(\hbar\omega_p\ \beta v) + 1$

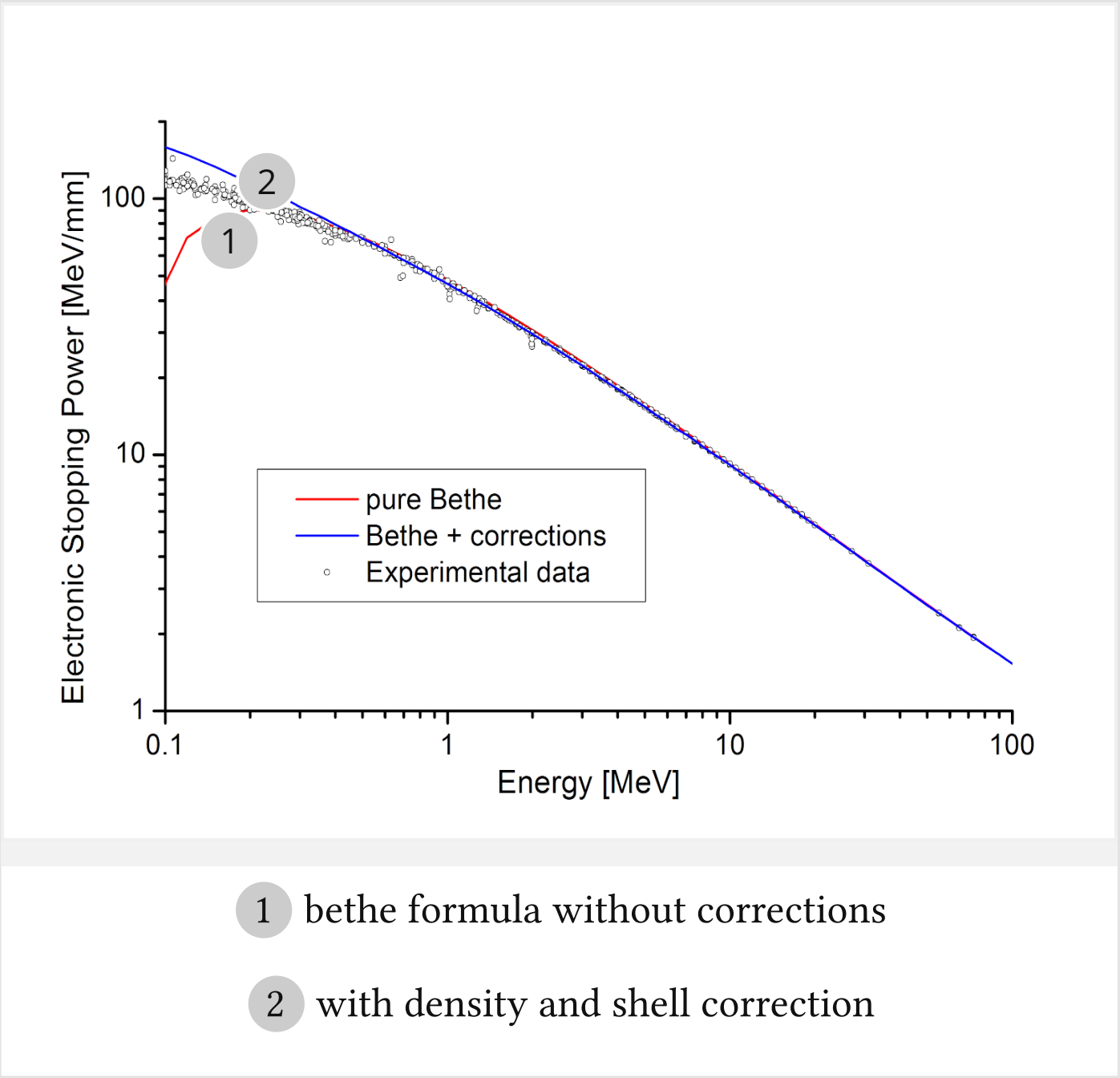

Fig 6 stopping power(energy loss) in Al for proton, showing the experimental data with Bethe in red and with corrections of density and shell in blue.

<u>Equation xxii</u> predicts the energy loss of particle passing through medium upto high accuracy from energy range of keV's to MeV's.

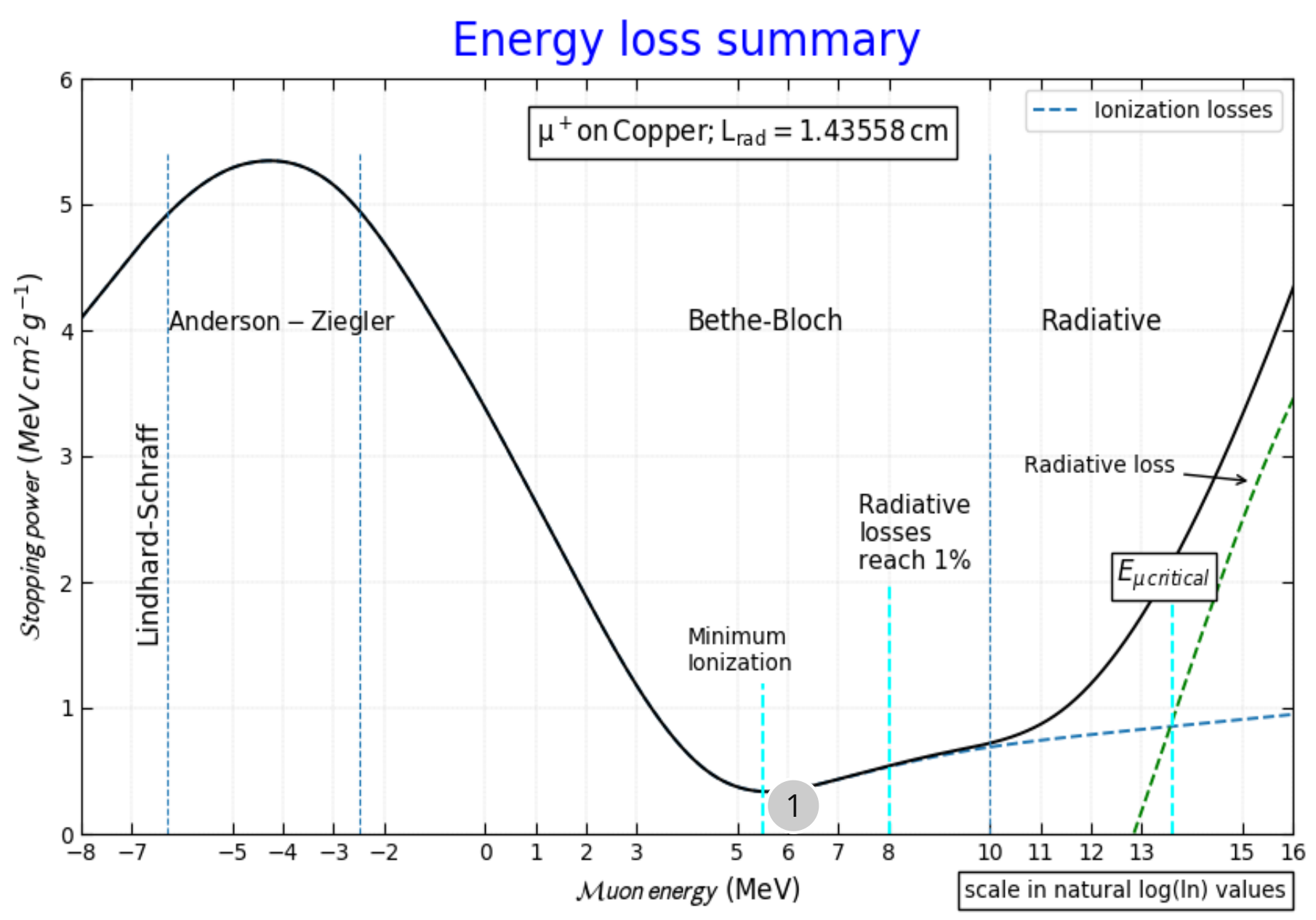

1 minimum energy loss

Fig 7 GEANT4 simulated energy loss of muon+ in Copper. Energy ranging from 300eV to 10 TeV's. Bethe-Bloch relation with corrections explains the curve from -8 to 10 MeV and high energies radiative losses contributes after 10 MeV's.

# Scaling of Bethe-Bloch formula

Equation xxii can be scaled to simpler forms under two given situations

1. Same particle incident on different target materials.
2. Different particles incident on same material.

For the first situtation; having same particle incidnet on different target materials, Bethe-Bloch relation can be approximately scaled as

$$\frac{1}{\rho}\frac{dE}{dx} = \frac{Z}{A}\, f(\beta, I) \qquad \text{( xxiii )}$$

For particle with high incident energy and for material(element) to be in the same period in the period table Equation xxiii can be approximated to a constant.

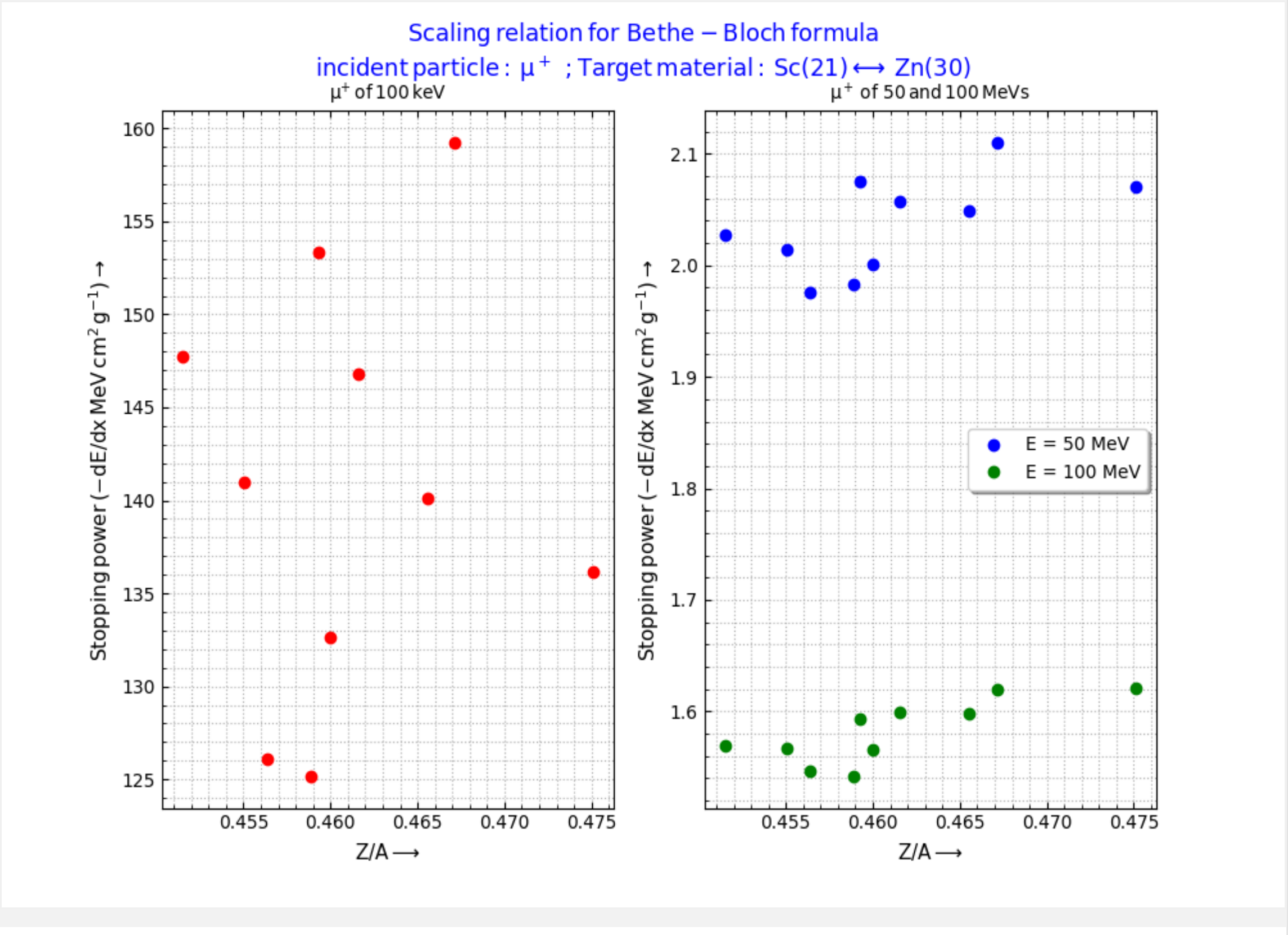

Fig 8 GEANT4 simulated data for muon+ incident on elements of first row of the d-block in the periodic table. Plotted the mass stopping power with ratio of Z/A, to approximate the function 'f' to a straight line or constant. It can be seen that mass stopping power is nearly a constant for higher incident energies as given in green and blue. Red shows the plot of muon+ for 100 keV energy.

For second situation; having different particles incident on same target material; Bethe-Bloch relations is scaled as

$$-\frac{dE}{dx} = z^2 f(\beta) \qquad \text{( xxiv )}$$

Equation xxiv can also be represented as

$$-\frac{dE}{dx} = z^2 g\left(\frac{E_{kin}}{M}\right) \ldots M{=}mass\ of\ the\ incindent\ particle \qquad \text{( xxv )}$$

# Range of charged particles

Equation xi can be modified in terms of energy to calculate the range of charged particles. The modified relation is given below

$$\mathcal{R} = \int_0^E \left(\frac{dE}{dx}\right)^{-1} dE \tag{xxvi}$$

Using Equation xxiv for representing the energy loss in terms of incident energy, range can be approximated as

$$\mathcal{R} = \int E^b \; dE \propto E^{1.75} \tag{xxvii}$$

Beacuse of the statistical nature of the range of charged particle, a semi-empirical relation is used instead to calculate the range given as

$$\mathcal{R} = \mathcal{R}_0 + a \int_{E_{min}}^{E_{max}} \left(\frac{dE}{dx}\right)^{-1} \; dE \ldots {}_{a=constant} \tag{xxviii}$$

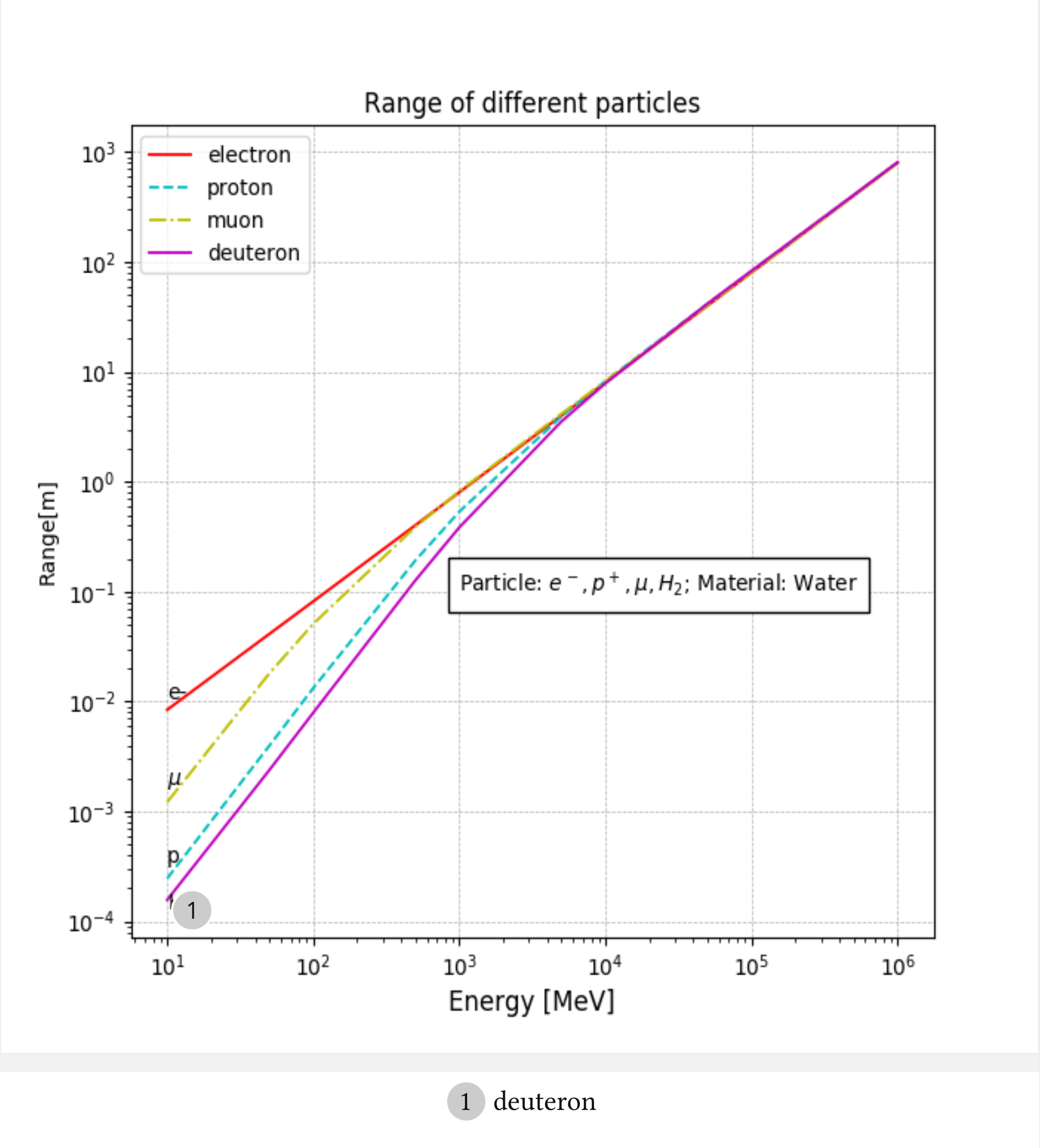

Fig 9 GEANT4 simualted data of range of different particles in water with energy range of 10MeV to 1 TeV.

For particle incident on different targets, range is approximated using *Bragg-Kleeman relation* given by $\mathcal{R}\rho \propto \sqrt{A}$.

### 2.2.3 Energy loss relation for electron and positron

Bethe-Bloch formula is that it is valid for heavier compared to electron such as *protons, muons, Kaons, Pions etc.* Although, the basic idea of collision loss is also valid for electrons and positrons, it is not able to explain energy loss due to electrons and positrons because of the given reasons

- The small mass of electron and positrons.
- The indistinguishability of incindent electrons with the electrons of the material.
- $E_{max}$ for this case becomes $\frac{E_{kin}}{2}$ for electron and positron.

The modified energy loss formula then becomes

$$-\langle \frac{dE}{dx} \rangle_e = 2\pi N_a r_e^2 m_e c^2 \rho \frac{Zz^2}{A\beta^2} \; [ln(\frac{\tau^2(\tau+2)m_e^2c^4}{2I^2}) + F(\tau) - \delta - \frac{C}{Z}] \ldots z{=}1 \qquad \text{( xxix )}$$

where,

$$F(\tau) = \begin{cases} 1 - \beta^2 + \frac{\frac{\tau^2}{8} - (2r+1)ln2}{(\tau+1)^2} & ; \quad e^- \\ 2ln2 - \frac{\beta^2}{12}(23 + \frac{14}{\tau+12} + \frac{10}{(\tau+2)^2} + \frac{4}{(\tau+3)^3}) & ; \quad e^+ \end{cases} \qquad \text{( xxx )}$$

if we ignore Equation xxx, then the above relation is similar to the Bethe-Bloch relation. Using Equation xxviii and substituting with the above relation, we can calculate the range of electrons and positrons.

It is important to note that for electrons and positrons of similar or same incident energy, the energy loss difference only comes from $F(\tau)$. Within the Bethe-Bloch range, this difference is not significant enough, therefore it can be taken to be approximately the same. The difference is contributed by the probability that positron may result into photon via *annihilation.*

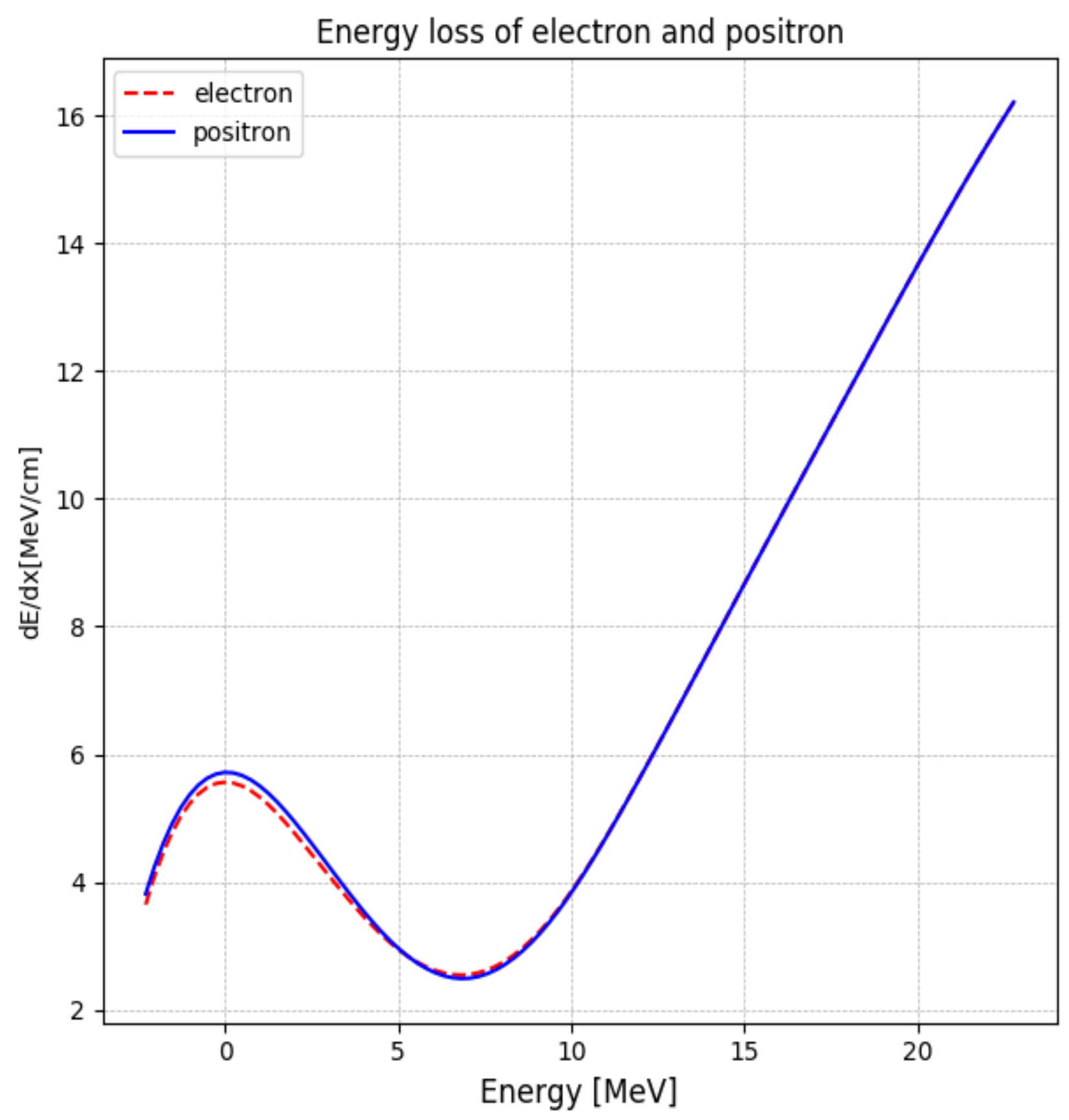

Fig 10 plot shows the stopping power (energy loss) of electrons(red) and postirons(blue) in lead with incident energy from keV's to TeV's. The energy axis marks the values as ln(E), (in the natural log scale) from 0 to 20. Used GEANT4 for simualting the energy loss in lead

## 2.3 Summary of Literature

To understand about how the particles lose energy while moving through matter required to understand some important concepts and definations which is essential for understanding of the energy loss processes. We learnt about the classical approach to develop the theory, its success and its failure. Then we studied about how Bethe solved the problem of the arbitrariness of impact parameter that resulted into the QM energy loss theory and relation. scaled the relation to simpler forms under consideration of particles and target materials. Understood the problems Bethe-Bloch relation contained for explaning the energy loss of electron and positrons and saw that calculation gave a new but similar relation. The calculations for some parameters will be given in appendix.

# 3 METHODOLOGY

## 3.1 GEANT4

GEANT4 is a simulating toolkit that was used to simulate the passage of particles through various materials. It is a standard simulating kit that is being used in various fields that require the simulation of particles through materials. The physics part is very vast in nautre and contains number of different processes and a wide range of particles and materials that could be used for simulation. This is majorly used with *Linux, Ubuntu* interfaces. In this we worked on Ubuntu 18.04 LTS and used GEANT4 for simualting and generation of data.

## 3.2 Methods

### 3.2.1 GEANT4 and OpenGL

In this study, I used the *physics module* and in that used the *electromagnetic sub-module* and in that used *TestEm0* module was used for simulating the passage of particles through different materials. The generation and output of data was executed as shown.

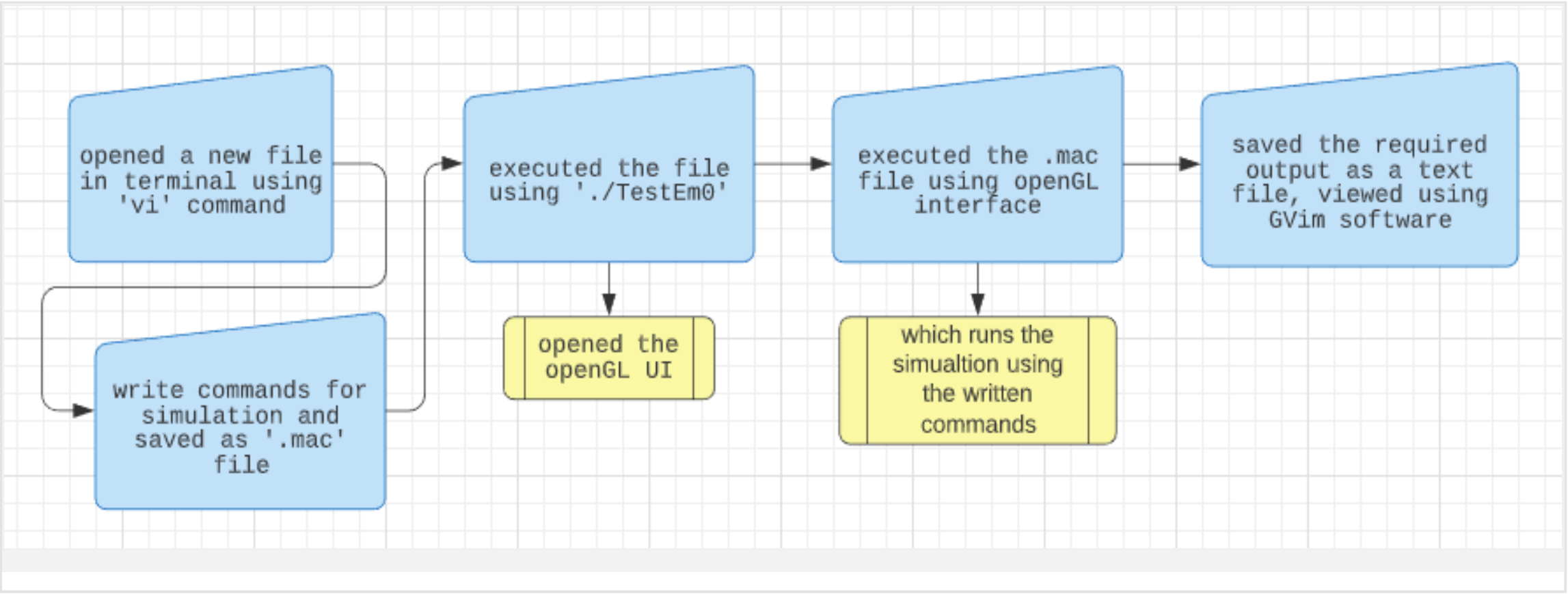

Flow chart 1 shows the process of generating using GEANT4 and storing the data as a text file.

The real data, for comparing the generated data, was taken from **NIST**(National Institute of Standards and Technology) **Reference Database 124.**

### 3.2.2 Python 3

For analysing or plotting the generated output from GEANT4, i used *Python 3.6.9* and *Sublime Text* as editor. The data has been plotted mainly using the matplotlib.pyplot library along with two other libraries *numpy, scipy.*

For few plots used the *Interplote* module of the *Scipy* library, for making the interpolation using *Spline method*. The following command has been used with change in the name of data file only

```
from scipy import interpolate
h = (max(energy)-min(energy))/1000.0
tck1 = interpolate.splrep(energy , electron , s=0)
energy_new = np.arange(min(energy) , max(energy) , h)
e_new = interpolate.splev(energy_new , tck1 , der=0)
ax1.plot(energy_new , e_new , 'b-' , label='e-')
```

Code 1 first statement is the importing of the interpolate module from scipy library. 2nd statement is dividing the energy domain into 1000 points. We used s=0 for no smoothing between two data points.

The fomr of the basic command line used for plotting is given below

```
import numpy as np
import matplotlib.pyplot as plt
from matplotlib.font_manager import FontProperties
from scipy import interpolate
from matplotlib.ticker import AutoMinorLocator, FormatStrFormatter
from scipy.optimize import curve_fit

a = (np.loadtxt('energyloss_all_copper.data'))##file name is changed for different data files
energy = a[: , 0]##variables changed for different plots
electron = a[: , 1]
```

```
proton = a[: , 2]
###using spline interpolation method
h = (max(energy)-min(energy))/1000.0
tck1 = interpolate.splrep(energy , electron , s=5)
energy_new = np.arange(min(energy) , max(energy) , h)
e_new = interpolate.splev(energy_new , tck1 , der=0)
###
fig , ax1 = plt.subplots(1 , 1 , figsize=(9 , 9))
fig.suptitle(r'$\mathcal{Stopping\/power\/of\/particles\/in
\/Copper}$' , fontsize=17)
ax1.plot(energy_new , e_new , 'b-' , label='e-')
ax1....
....
plt.show()
```

Code 2 this shows the basic structure of the commands used for plotting the data.

# 4 RESULTS AND DISCUSSION

## 4.1 Results for alpha particle incident on Aluminium and Copper

### Alpha incident on Aluminium

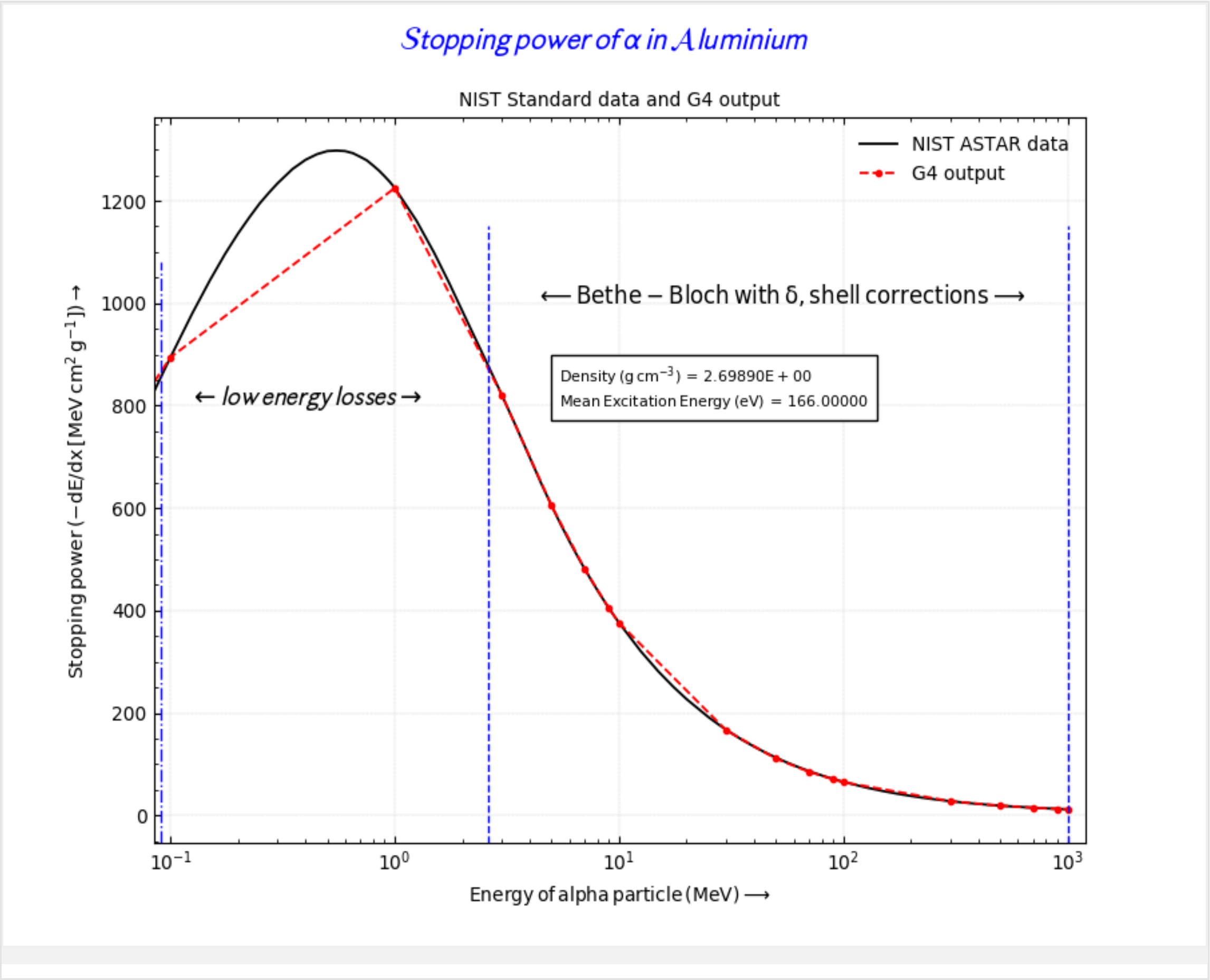

Figure 1 The red plot shows the generated output of GEANT4 simulation of alpha particle incident on Aluminium. NIST data has been represented for the same in black.

## Alpha incident on Copper

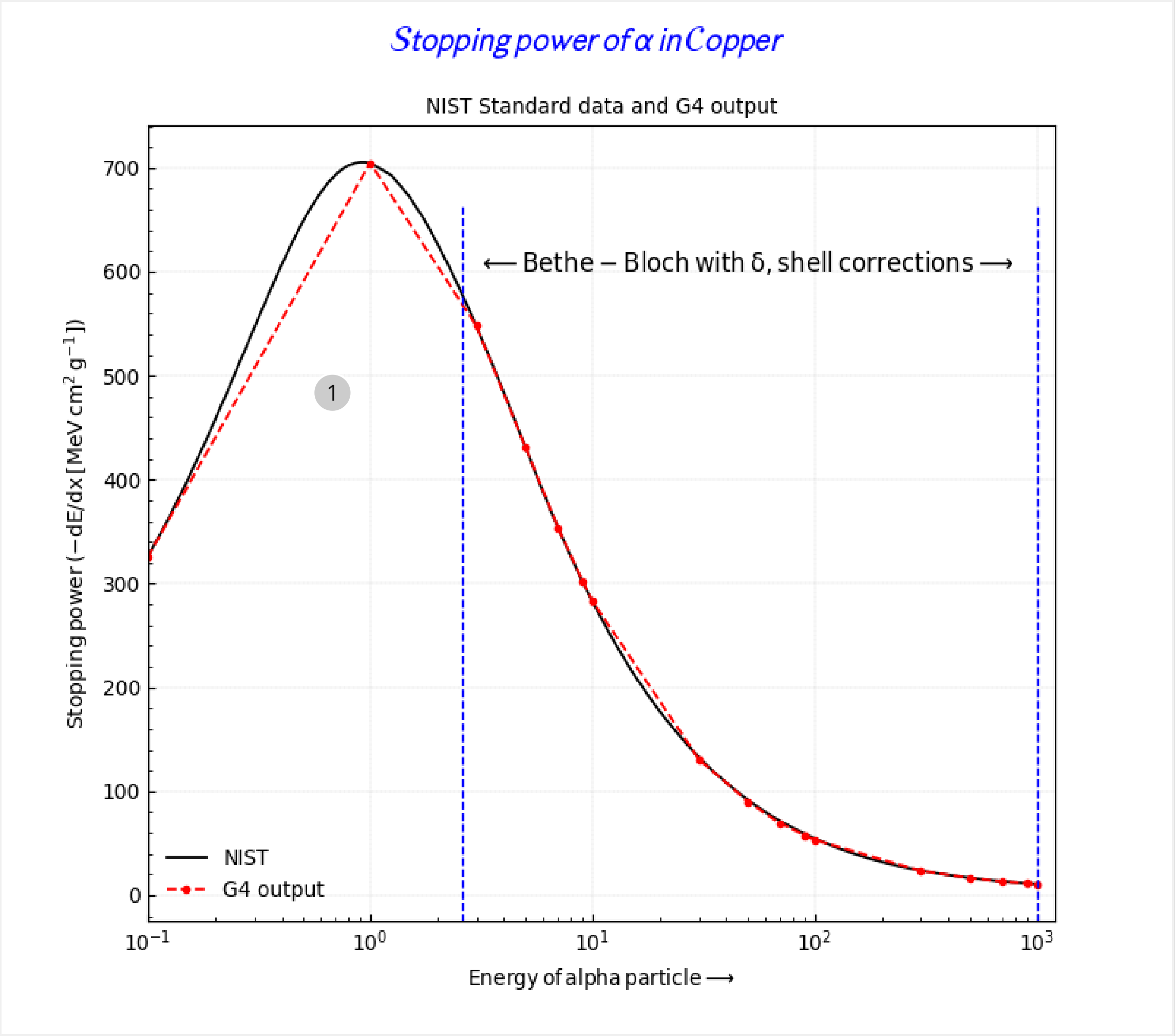

1 low energy loss region

Figure 2 plot shows the energy loss of the GEANT4 simulated data (in red) and NIST data (in black). Energy is in MeV's.

We can see that in both the plots the GEANT4 simulated data overlaps to the NIST data to a great extent in the energy range of *1 to 1e3 MeV.* In the low energy region, the energy data points were taken at certain seperated points. This plot has been plotted using *Linear interpolation.*

## Comparison of the two plots

1. The maximum energy loss of alpha occur for incident energy to be near 1 MeV with higher in Aluminium.
2. The rate of energy loss of alpha particle is more for Aluminium than Copper for energy range of 1 MeV to 10 MeV.
3. Both the energy loss reaches nearly the same value i.e nearly 0 (but will be more than 0) for 1e3 MeV energy alpha particle.

## 4.2 Results for proton incident on Water and Copper

### Proton incident on Water

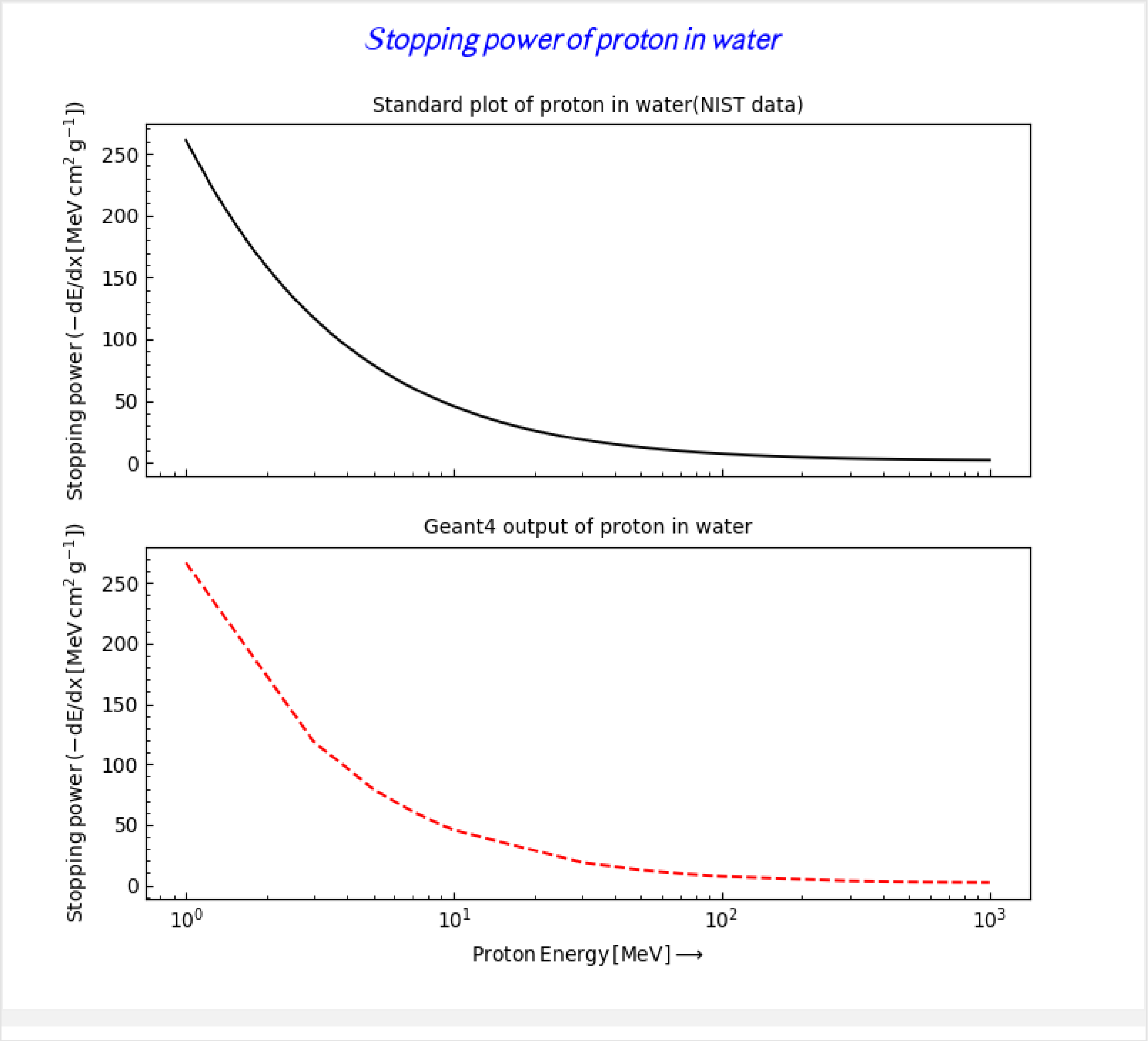

Figure 3

the plot shows the variation of the mass stopping power of proton plotted with incident energy in MeV in water(liquid). the upper part in black shows the NIST data and lower part in red shows the GEANT4 simulated data

## Proton incident on Copper

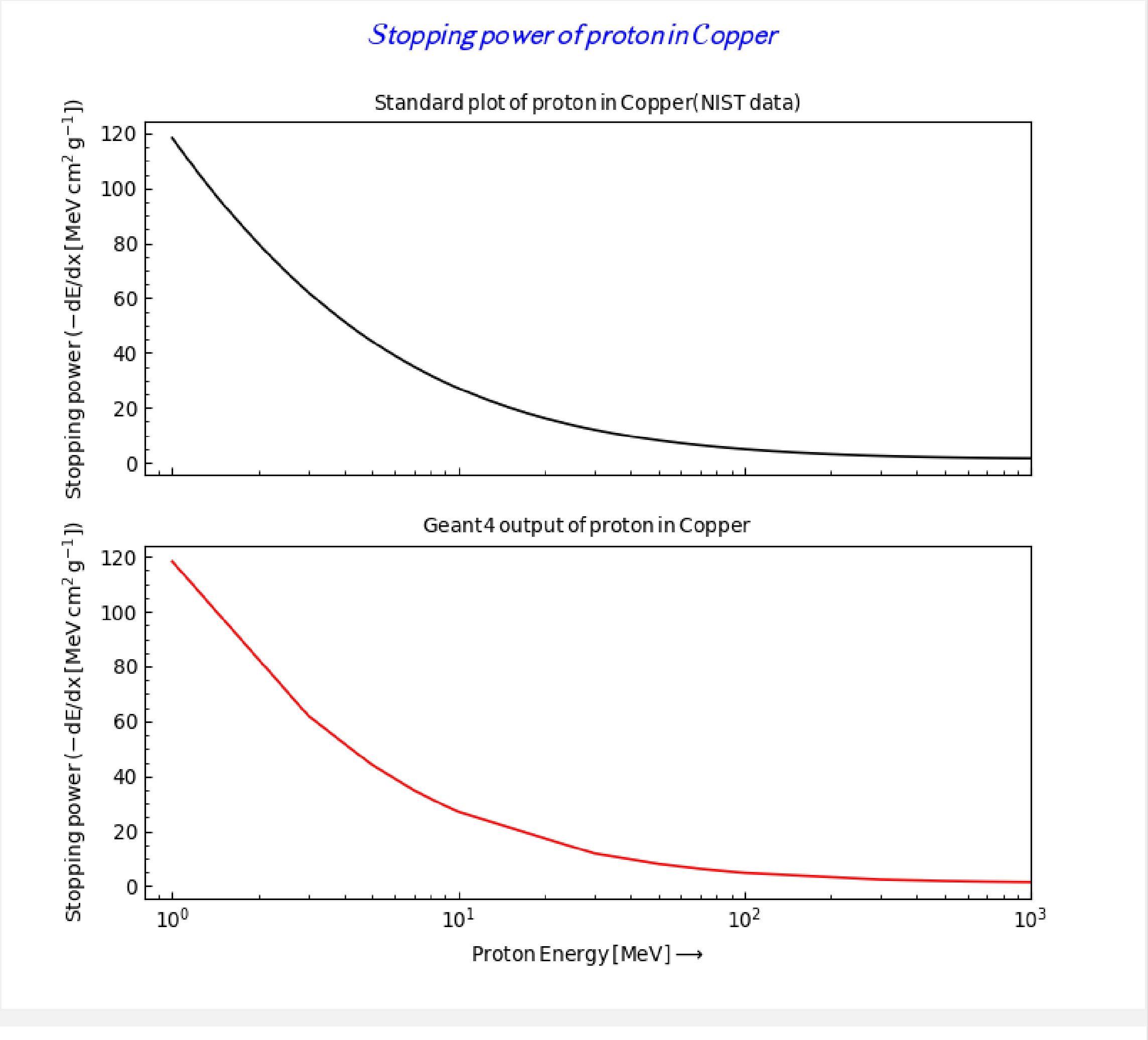

Figure 4 the plot shows the variation of the mass stopping power of proton plotted with incident energy in MeV in copper. the upper part in black shows the NIST data and lower part in red shows the GEANT4 simulated data

We can see that in both the plots, the GEANT4 simulated data; shows the same variation; to a high extent as that of the NIST data. We can see that with increase in incident energy, the mass stopping power decreases to a minimum in both the plots.

## Comparision of plots

1. The energy loss for proton in water is more than that in Copper for entire energy range.
2. The amount of energy lost by electron in Water is nearly *twice* that is lost in Copper.

# 4.3 Results for electron incident on Water and Copper

## Electron incident on Water

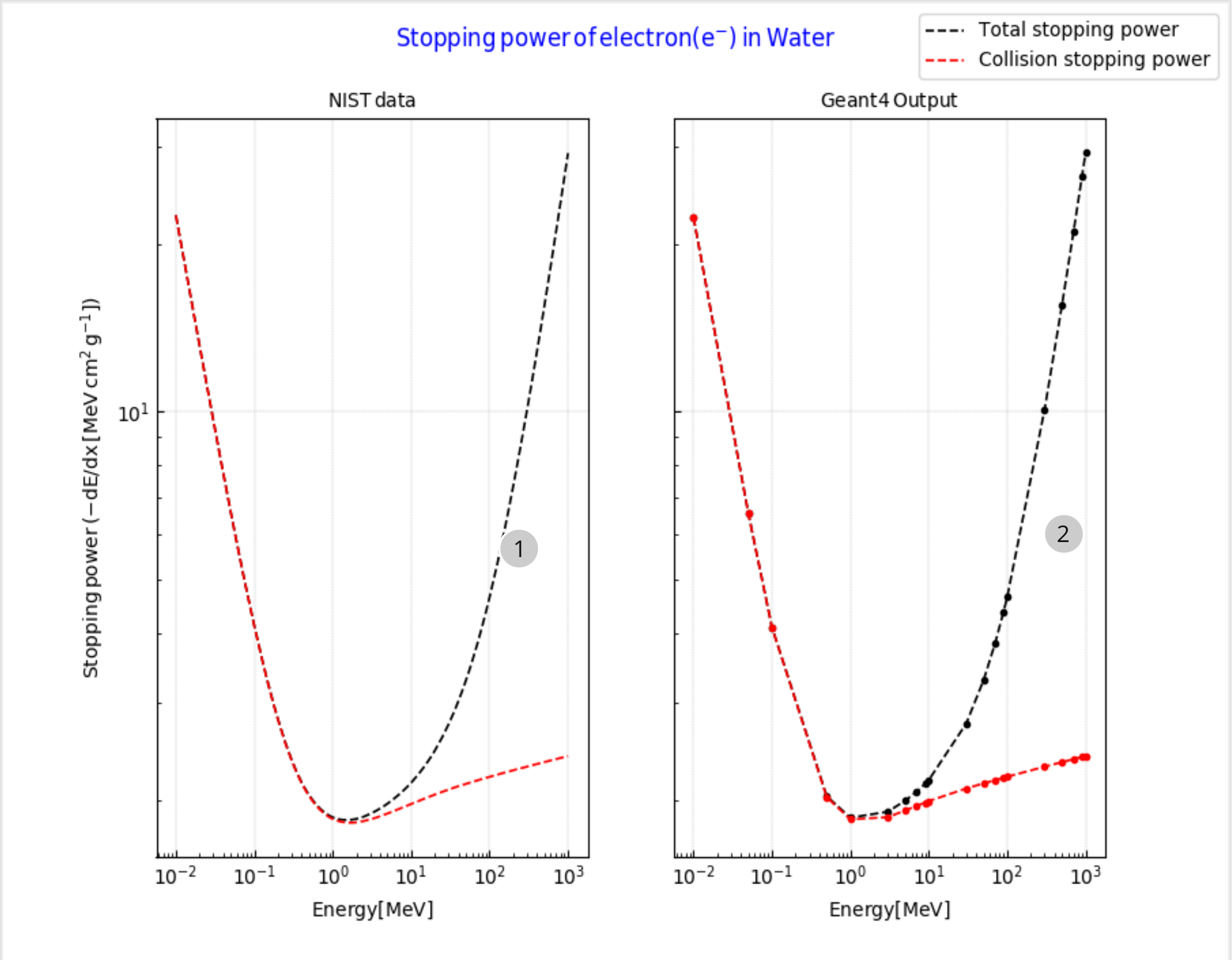

1 radiative loss 2 radiative loss

Figure 5 Plot shows the variation of mass stopping power of electrons in Water(liquid), the left shows the NIST data while right shows the GEANT4 simulated output. Both axes in 'log' scale.

## Electron incident on Copper

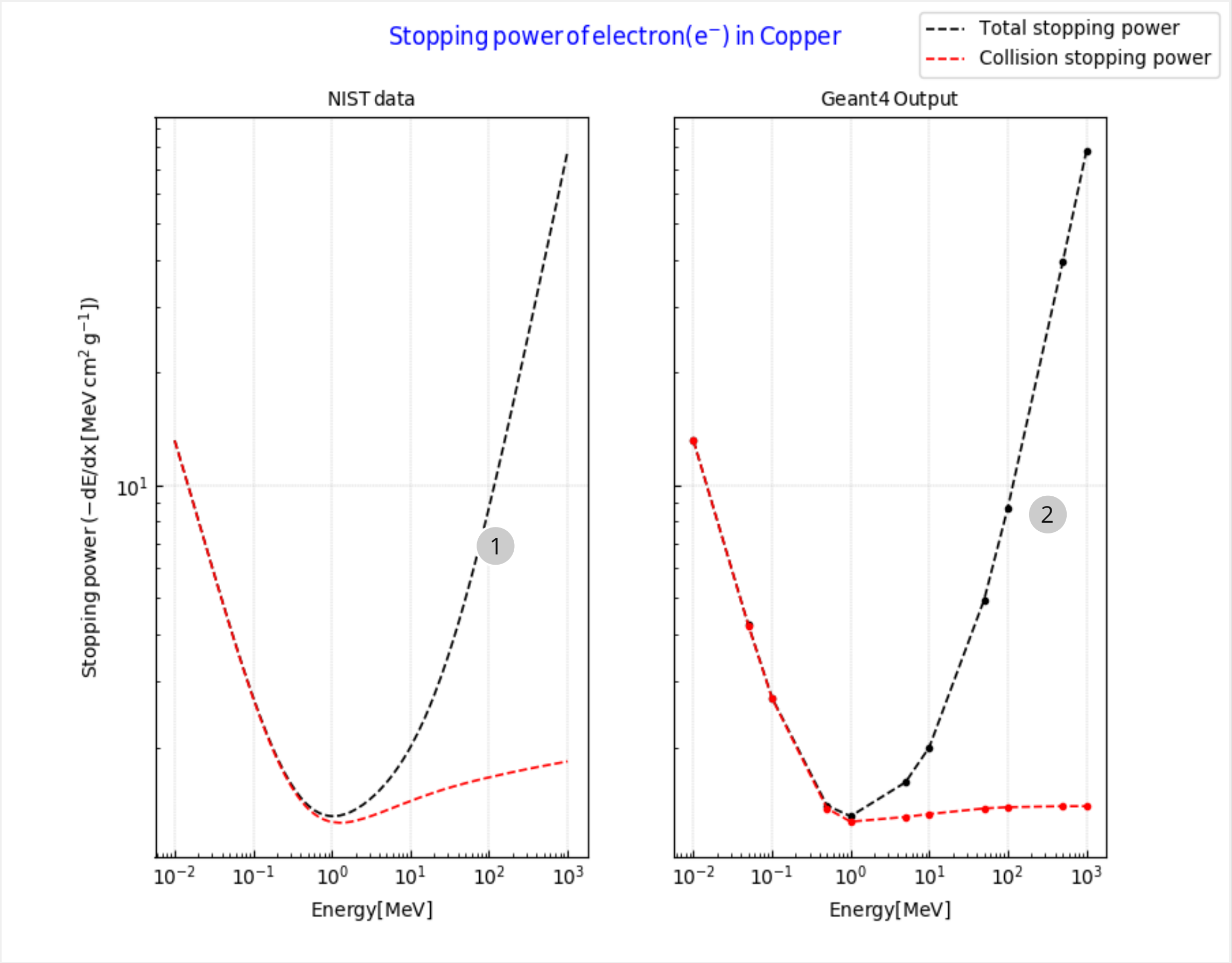

1 radiative loss 2 radiative loss

Figure 6 Plot shows the variation of mass stopping power of electrons (both for collision and total losses) in Copper, the left plot shows the NIST data, right shows the GEANT4 simulated output. Both axes in the 'log' scale.

In the case of electron, plot shows both the collision loss and the total energy loss, which also includes the radiatve losses at high energies. The increase in the total energy loss is contributed mainly by *Bremsstrahlung Radiative loss.*

## Observations from the plots

1. The collision stopping power, in both plots, shows similar variation to NIST data; with a slight more observable *relativistic rise* for NIST data than GEANT4 data in Figure 5.
2. The minimum in the energy loss occurs at 1 MeV for Copper and 1 MeV (approx.) for Water.
3. The rise in collision loss is due to corrections (δ and C) in Bethe-Bloch formula.

4. The stopping power or energy loss by electron is more in Water than Copper.

## Problem faced with 'log' scale for plotting the energy loss of electron.

The use of log scale in the electron case, hindered from estimating the quantitative difference in the energy loss by the electrons in the two materials.

# 5 CONCLUSION

Q1. What is the *conclusion* of the results?

From the results above, we can conclude that Beth-Bloch relation explains the energy loss via collisions for charged particle moving through matter and it has been verified by using GEANT4 simualtion data and comparing it with the Standard NIST data.

Q2. What *I learned* from this study?

Learned about the theory behind the observational phenomenon of energy loss of particles in matter. About the different interactions throught which a particle may interact with the matter. Learned about the historical developments that led to the birth of the energy loss theory and how it was modifed using different observations and theories. In the practical side, learned to use GEANT4 toolkit for making simulations, how to write the simulation code and execute and how to use OpenGL interface.

## Q3. Why this study is *important*?

*Firstly* to gain knowledge about the nature of fundamental particles that makes this *universe.* The interaction of a particle gives a lot of information abut the particle. *Secondly,* to develop the technology that is based upon particles such as *proton thearpy, FNT (fast neutron therapy), making radiative sheilds, protective gears for astronauts, nuclear scientists* etc. understanding of particle ineraction with matter is of crucial importance.

## Q4. What are the short-comings of this study?

In this study, we only studied about the interaction of charged particle with matter. We didn't studied about the ineractions of neutral particles(neutrons), photons (X-rays, gamma-rays) etc. We didnot stuided about the low energy losses that is responsible for the *initial rise in the stopping power.*

# ACKNOWLEDGEMENTS

I would like to start by expressing my gratitude of thanks to the *Indian Academy of Sciences* for providing with the wounderful opportunity of SRFP fellowship, through which I was able to work under the guidance of guide from IISc, Bengaluru and learned a lot of new things and gained new skills. I would also like to give a special thanks to the academy and to the people, for organizing this year's fellowship in the time of Covid-19 pandemic. I would like to express my deepest thanks to my mentor Dr. Jyothsna mam for providing the guidance and . Along the same, I would also like to thanks Lata Panwar mam, who helped me in solving the queries I had regarding the project or techanical queries with the software and helped me in understanding the software. I would also like to express my gratitude to my fellow projectmate, Pradeeptha R Jain from Mangalore, Karnataka, who helped me in solving queries related to both techanical and reading related, for also discussing the small-small issues whenever I had. Lastly, I would like to thank my sister, who helped me in writing the report and parents who bought a new internet connection upon my request.

# APPENDICES

# A : Density and Shell Corrections

## A.1 Density corrections(δ)

The density effect is due to the polarising property of the charged particle. It results in the reduction of energy loss than expected. This arises because of the higher shielding of the far away electrons by the near ones to the incident particle, which shield the effective interaction of the particle with the far away electrons. Therefore the collision energy loss gets reduced by an effective amount of δ.

The variation of this δ with the incident particle energy and the fundamental frequency of the interacting matter is given as

$$\delta(\beta\gamma) = ln(\frac{\hbar\omega_p}{I}\ \beta\gamma) + 1$$

an emprical relation of δ was given by Steinheimer as

$$\delta = \begin{cases} 0 & ; \quad X \leq X_0 \\ 4.6052X + C + a(X_1 - X)^m & ; \quad X_0 < X < X_1 \\ 4.6052 + C & ; \quad X > X_1 \end{cases}$$

where X = log(βγ) and $X_0, X_1$, C , a , m depends upon the absorber and the form of C is given by

$$C = -(2ln(\frac{I}{\hbar\nu_p}) + 1), \textit{where } \nu_p = \sqrt{\frac{Ne^2}{\pi m_e}}; N_e = \frac{N_a Z\rho}{A}$$

where $\nu_p$ is known as the plasma frequency.

### A.2 Shell Corrections (C)

The shell correction amounts for the assumption taken in deriving the classical and Bethe relation, that the electron is assumed to be stationary. This assumption breaks down if the velocity of the incident particle becomes comparable to the electrons orbital velocity.

## B: Adiabiatic Invariance

This assumption states that the energy transfer from the incident particle to that electron, taken as orbiting, should take place with a time period which is less than the orbital time-period of the electron. If the time is more than the orbital time period than the electron energy will not be changes.

This could be understood as the overlapping of two waves, one of the electron orbiting the nucleus and the other of the energy carrying wave to the electron. If we want that the final energy of the electron should represent the transfered energy, the frequency of the wave carrying the energy to be tranferred should have a high frequency or small time period, so that the effect of the incident wave is clearly reflected by their superposition.

## C: High energy losses

### C.1 Bremsstrahlung Radiation loss

Whenever a moving charged particle is deflected or deaccerlated by the Coulomb field of nucleus, the loss in the kinetic energy is conserved by the emmision of the photons at that point. This radiation is called *Bremsstrahlung Radiation.*

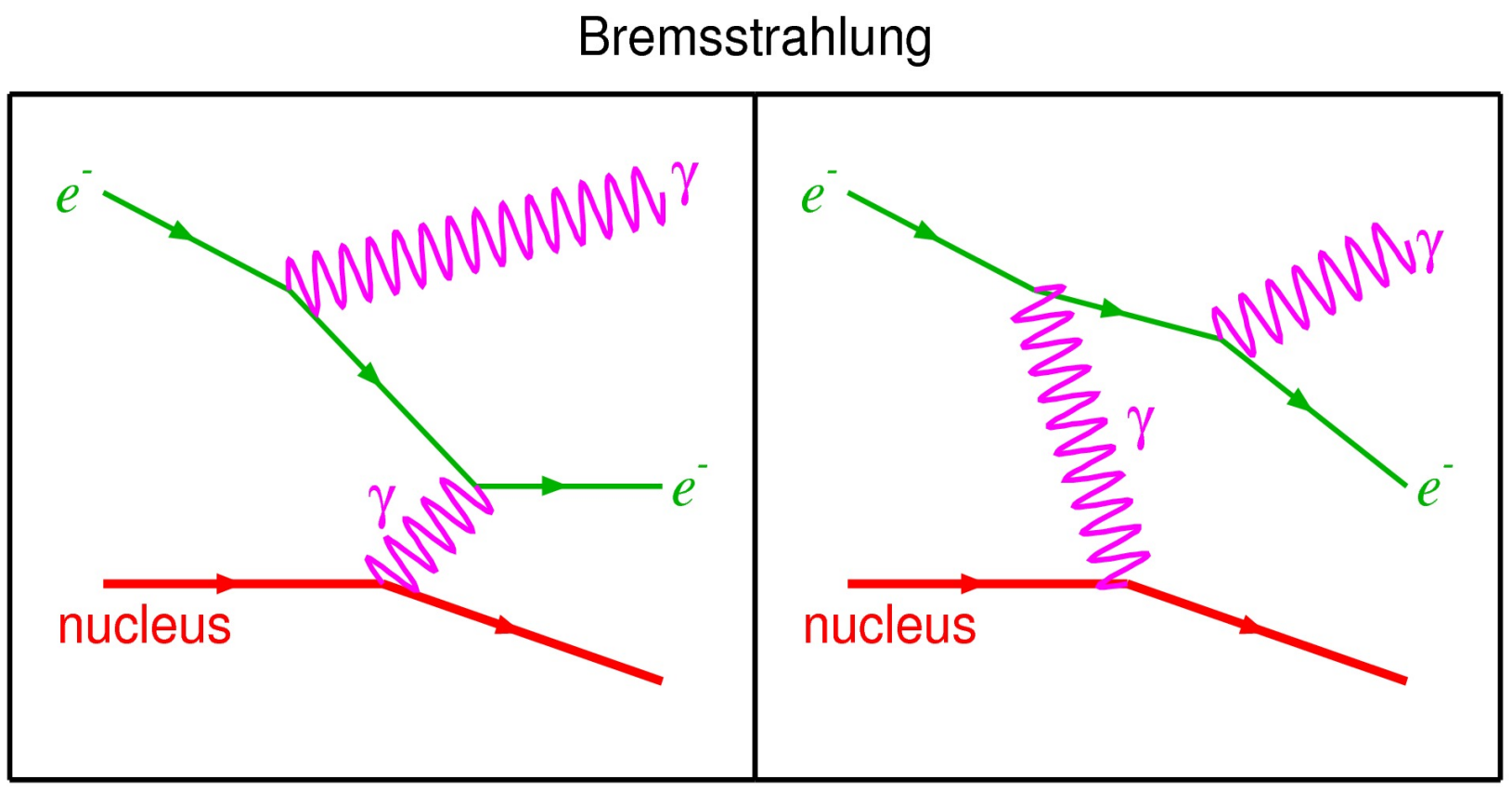

feynmman diagram of production of bremsstrahlung radiation for electron in the presence of nuclues. The photon interacting mediating the electron and nucleus is a virtual photon showing the interaction. $\frac{e^2}{4\pi\epsilon_0 mc^2}$

The energy loss through Bremsstrahlung Radiation for incident particle with high energies is given by

$$-\frac{dE}{dx} = 4\alpha N_a \frac{Z^2}{A} z^2 r^2 E\ ln(\frac{183}{Z^{1/3}})$$

where E is the energy of the incident particle and r is given by $\frac{e^2}{4\pi\epsilon_0 mc^2}$, with 'm' being the mass of the particle.

## C2. Cherenkov Radiation

The energy loss by the particle in the form of radiation whenever the speed of particle exceeds the phase velocity of the wave in a dielectric medium. This gives a mathematical condtion on β depending upon the refractive index η as

$$v_{incident\ particle} > \frac{c}{\eta} \implies \frac{1}{\eta} \leq \beta < 1 \ldots for\ a\ particle\ with\ mass$$

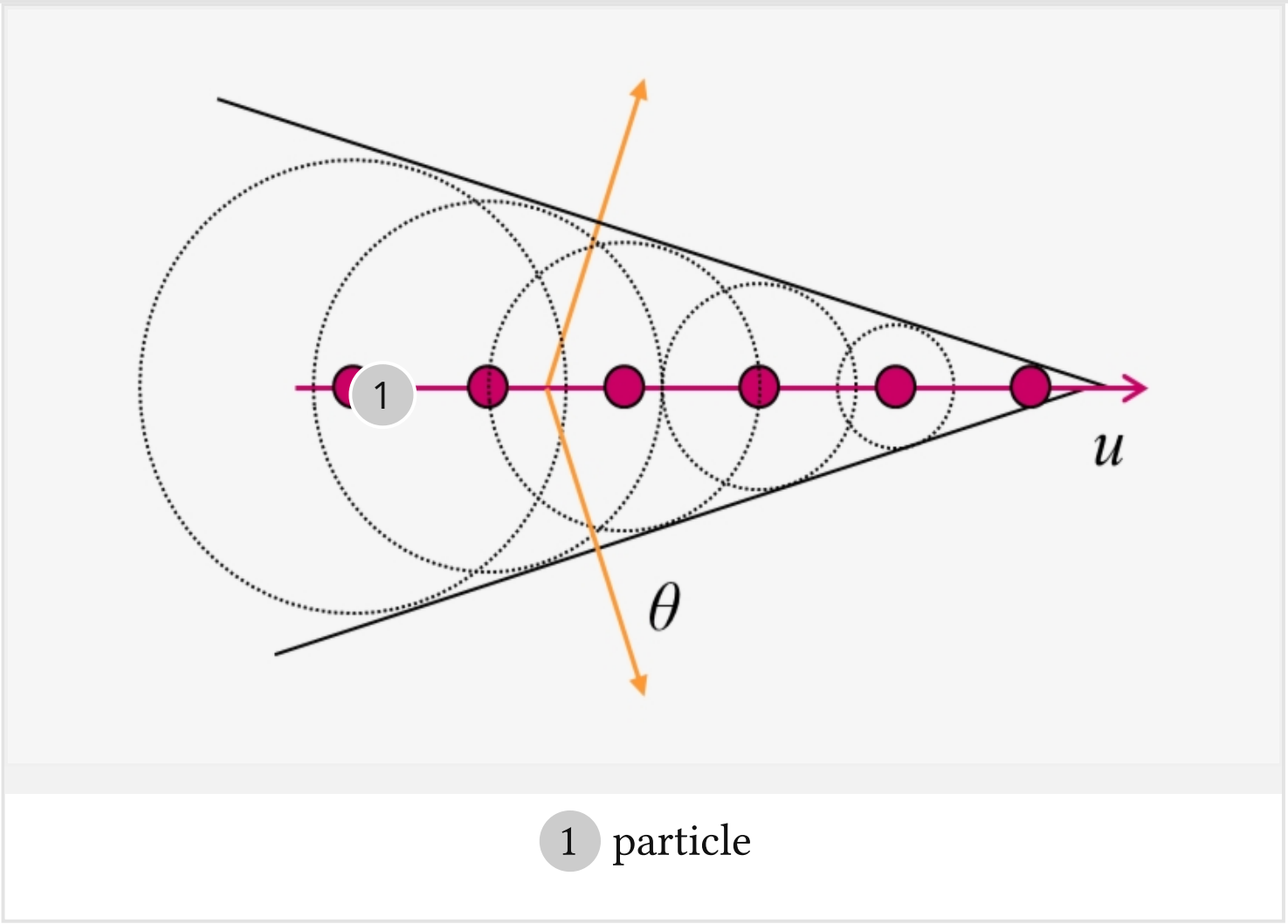

Shows the light cone formed by the particle moving at speed u emitting the cherenkov radiatons, showed by the growing circles with the direction of emission with arrows.

The energy loss by the incident particle in the form of cherenkov radiation is given as

$$\frac{dE}{dx} = 2\pi hc\alpha \int_{\lambda_1}^{\lambda_2} \left(1 - \frac{1}{\beta^2 \eta^2}\right) \frac{1}{\lambda^3}\, d\lambda$$

where the refractive constant of the material, 'η' is dependent upon the wavelength λ.

# Source